# The Biological Consequences of the Computational World: Mathematical Reflections on Cancer Biology[1]

Giuseppe Longo
Centre Cavaillès, République des Savoirs,
CNRS, Collège de France et Ecole Normale Supérieure, Paris,
and Department of Integrative Physiology and Pathobiology,
Tufts University School of Medicine, Boston
http://www.di.ens.fr/users/longo

**Abstract** The role of continua has been clear since antiquity in the mathematical approaches to physics, while discrete manifolds were brought to the limelight mostly by Quantum and Information Theories, in the XX century. We first recall how theorizing and measuring radically change in physics when using discrete vs. continuous mathematical manifolds. It will follow that the reference to discrete structures and digital information is far from neutral in knowledge construction. In biology, in particular, the introduction of information as a new observable on discrete data types has been promoting a dramatic reorganization of the tools for knowledge. We briefly analyze the origin and the nature, then some consequences of the bias thus induced in life sciences, with particular emphasis on research on cancer. We finally summarize new theoretical frames that propose different directions as for the organizing principles for biological thinking and experimenting, including in cancer research. Cancer is now viewed as an organismal, tissue based issue, according to the perspective proposed in (Sonnenschein, Soto, 1999; Baker, 2015).

## 1. Introduction: discrete vs. continuous manifolds

The computational virtuality is heavily affecting common and even scientific knowledge. The new symbolic forms of interaction on electronic digital networks provide extraordinary new tools for mankind, from everyday worldwide exchanges to fantastic scientific modeling. They also suggest an image of the world rich of a peculiar bias. It is in biology that the reference to informational, alphanumerical data structures has had the greatest impact throughout the second half of the twentieth century, by making DNA an "information carrier" or even a "computer program" for ontogenesis. As a consequence, development has been interpreted as the deployment of a program and organisms as "avatars" of genetic information[2]. We will mention some of the strong

---

1 This paper has been made possible by many years of a very stimulating collaboration with C. Sonnenschein and A.M. Soto, biologists of cancer at Tufts University. The third part of G. Longo, "Le conseguenze della filosofia" i*n* **"A Plea for Balance in Philosophy"**, R. Lanfredini ed., ETS, Pisa, 2015, is a very preliminary version of this text (a reduced translation of that paper is in  http://www.glass-bead.org/article/the-consequences-of-philosophy/?lang=enview )

2 **On Avatars.** From (Gouyon et al., 2002; pp. 154-5), a well-known text book on neo-Darwinian Evolution "To denote that which transmits genetic information or its physical carrier, we use the term avatar borrowed from the Hindu religion; it alludes to the physical forms adopted by the god Vishnu on his visits to Earth … The avatar, as noted by J. Damuth, interacts with the environment which provides for its needs and exerts an influence upon it but, consequences of this weak conceptual frame based on a vague, common sense reference to computational notions[3], with particular emphasis on cancer research.

It should be clear that Information Sciences as such do not imply, per se, the fuzzy applications of their notions to other domains. Yet, they contain the grounds for a reading of the world through the digital or "discrete" grid of numerical databases and computations, as soon as those fantastic tools for digital computing are transformed in "models" or *true images* of physical or biological phenomena. In other words, we claim that the intelligibility of the world proposed by discrete mathematical structures is far from neutral: typically, it yields a peculiar approach to causality.

**1.1 Physical causality and discrete manifolds**
By "discrete" here we refer to the only good mathematical sense one can give to this notion: the points of a discrete manifold can be "naturally" given the discrete topology, that is, they may be all isolated, each in its own neighborhood[4]. B. Riemann (1854) beautifully expressed this in his fundamental writing that opened the way to differential Geometry and, then, Relativity Theory: « … in a discrete manifold, the ground of its metric relations is given in the notion of it, while in a continuous manifold, this ground must come from outside. Therefore, either the reality which underlies space must form a discrete manifold, or we must seek the ground of its metric relations outside it, in binding forces which act upon it.» In other words, a discrete, complete manifold forces a unique metrics, where each point is "naturally" isolated and there is a minimal distance,[5] while the metrics (and, thus, the curvature of space that he correlated to the metrics by his fundamental result) is grounded on the *causal relations* (the "forces acting on it"). Einstein understands gravity as a cause of falling bodies, by identifying it to inertia in curbing Riemannian spaces[6].

James Jeans, a major (quantum) physicist of the early XX century, insists: "when discontinuity gets in, causality gets out". A discrete manifold is totally discontinuous or totally disconnected: its scattered points have no topological connection with each other. Note that Quantum Physics and its indeterminism are presented in space and time *continua*: "discrete" structures appear in the dimension of energy (or in the dimension of Planck's h, an action, i.e. energy × time). Typically, the energy spectrum of the bound electron is discrete, a true surprise in 1900, while the free electron has a continuous spectrum. As a special case of quantum indeterminism, 0-1 alternatives may also result at measurement, such as the spin-up or spin-down of an electron; then the "standard" interpretation consistently and audaciously claims that this event "has no causes", it is pure contingency – thus, "causality gets out". From Einstein to Böhm and De Broglie, some physicists rejected this interpretation and many still search for "hidden variables" or hidden causes varying in an underlying continuum. These scientists hoped that hidden causes (hidden variables in continua) could also justify quantum entanglement, that is, probability correlations in measurements of remote

---

above all, the avatar is produced by genetic information to ensure that this information is passed on. *Individual organisms easily meet this definition.* They interact with the environment, are produced by genetic information, and copy the information . .. . Selection targets only genetic information, *avatars are mere vehicles.*" [Italics added].

3  A. Danchin (2003; 2009) is one of the few biologists who tried to search rigorously for compilers and operating systems in DNA, while exploring even a possible genetic meaning of Gödel's theorem (see below for this peculiar case of "gödelitis").

4  If you consider the continuum of the real number line, the discrete topology on it is surely not "natural": all maps are continuous on it and no relevant mathematics can be done with this. The so called "natural topology" on the real line is usually considered the "interval topology" (or metrics); it is "natural" as it is derived from classical measurement in physics, which is always an interval (classical, and relativistic, measurement is approximated, it is given as a continuous interval, by principle – no jumps, no holes). "Naturality" can also be formally defined in fully general Category Theoretic terms, see (Asperti, Longo, 1991).

5  Typically, there are no accumulation points, that is limits points for a converging series, or operations converging to the limit make no sense.

6  **On Symmetries**. Along this idea, XXth century physics largely replaced causality by "symmetry properties", as extensively discussed in (Bailly, Longo, 2011). Indeed, by the unification of inertia and gravity, Einstein could say that a body falls for "symmetry reasons" (inertia is a symmetry property in the equations, by Noether's Theorems, see the reference). Yet, so far, in biology it may be wiser to preserve a "causal" terminology.

events (Jaeger, 2009). Note that this is yet another phenomenon that prevents from attributing to quantum space-time a discrete structure, as well separated small boxes of the size of Planck's length, say. By entanglement, quantum observables cannot be "separated" by measurement (there are instantaneous probability correlations, even at a distance). Thus, we are particularly far from the discrete topology, made of isolated, totally disconnected points.

In other words, discrete structures or discretized events provide an *a-causal* image of the world. The key issue of measurement, as the only form of access we have to phenomena, is set aside: both classical approximated measurement (an interval in continua) and the challenges of quantum measurement are forgotten (indetermination, entanglement). Digital databases are accessed exactly, a non trivial technological achievement in actual computers, and the causal relations are replaced by discrete dynamics of "information" encoded by digits; this dynamics follows formal rules or instructions on how changes of digits have to take place, that is following a "program". These replacement rules (replace a 0 by a 1, or vice versa) physically function according to hidden forces that act on discrete structures, that is, by varying on underlying continua. But, then, how does a digital computer actually work?

**1.2 Computational dynamics**
Modern computers are based on a fundamental idea by Turing (1936): namely, the split between software and hardware, as for the *elaboration* of information. The autonomous science of software or of programming was born, with its general mathematical frames in some fantastic areas of great mathematical rigor and achievements, thanks to Turing, Gödel, Church and a few others (Computability Theory, Proof Theory, Type Theory, thanks to which the author of these lines earned most of his living). The core idea is that programming and its science is independent from the hardware[7]. Similar conclusions can be drawn from Shannon's theory of *transmission* of information (1948): its analysis is independent from the material structure for the transmission (cables, waves, drums …).

Thus, programming may be identified with a general form of "term (re-)writing": programs are an alpha-numeric writing of instructions on how to transform or re-write alphanumeric strings into new alphanumeric strings (Bezem et al., 2003). In computer networks, distributed in space-time continua, this presents some peculiar difficulties adequately dealt by the difficult mathematics of concurrent and network programming. This is based, when needed, on continuous dynamics of complex structures (Baccelli, 2016a; 2016b), yet, these dynamics are still grounded and act on discrete data bases by term re-writing (Aceto et al., 2003). However, if one looks closely into a computer's hardware, the instructions that modify a discrete data type act by variations of electric tension's levels *in continuous fields* and/or *by driving* electric currents *in continua* into two stable states, throughout discrete thresholds. That is, in silico, continuous dynamics undergo "critical transitions", such as switches, that stabilize current or no current states in a material component of the hardware (the 0 and 1 at the base of computing). So, physical causes still refer to continua, yet the physical structure of computers allows seeing only a discrete interface, "pixel by pixel", where causality is hidden and only the writing and re-writing system appears (the changing 0 and 1's). This is an amazing technological construction, whose discrete visible image is as far from the world as the invention of the alphabet, some 5.000 years ago in Mesopotamia (Herrenschmidt 2007). At that time, as paleo-anthropologists claim, humans first discretized the continuous flow of language, originally a song. Indeed, modern digital computers are the latest advancement of that atomistic invention of ours, which cuts the flow of language into meaningless letters. Today, the alphabetic writing, once static, *moves* on a screen, it is not only written, but it is *re-written* according to written instructions. So, causality gets out from the image of the world that is proposed by re-writing machines, as it is hidden by a cascade of major technological inventions. We only see pixels, re-written from other pixel, 0s transformed in 1s and vice versa, by exactly-defined instructions in a discrete structure, with no idea for the biologist nor relevance for the computer scientist of how this is physically obtained. This is a fantastic accomplishment by the science of software (Theory of Programming), which has been broadly developed, independently of the hardware support and its causal dynamics. Instead, the analysis of a causal structure may be relevant in the natural sciences (e.g. one searches for the "causes" of cancer), possibly even to exclude causes, as standard quantum physics dares to do (the a-causal nature of the spin-up or down of a quanton).

---

7   Note that a major difficulty in realizing, concretely, Quantum Computing is due to the constraints that the physical theory (thus, the hardware) imposes to programming and to the unavoidable blend of hardware and software: e.g. measurement, which co-constructs the quantum state, and entanglement have key programming (software) consequences. And these are major challenges well beyond current information theories and technologies.

And here is another fantastic feature of discrete computations and information technologies: any set of isolated points can be (isomorphically) encoded just in one dimension - a *sequence* of 0 and 1's suffices – discrete data and computations are insensitive to dimensional coding. This is essential in order to encode Turing's Universal Machine and, thus, today's operating systems and compilers: they are encoded like programs and data, all in the same, unique dimension, the "Type" (or dimension) of numbers. The expressiveness of computing is based on the self-referential power of recursion and compiling, all encodable in the Type of integer numbers, a fantastic invention by Gödel, Church, Kleene and Turing. But this feat has been obtained by its insensitivity to dimensions (and to codings, modulo some coding costs).

Once again, these are very effective tools, but may yield a totally distorted image of the physical and biological world[8]. Typically, everything changes in physics and, a fortiori, in biology as well when changing dimensions: from the dimensionality of energy vs. force, say, to the description of waves' propagation, heat for example, dimensional differences are crucial, in physics; in biology, if one forgets dimensionality then one may also miss the bodily material structure of organisms, which necessarily is three space dimensions[9].

In conclusion, the informational/computational approach diverts attention from the rich networks of causal relations, within an organism and an ecosystem, in favor of an instructional *a-causal* perspective. A change in a phenotype *must* derive from a change in the instructions that are encoded in discrete data types which may totally bypass the physico-chemical causal structure or even force a wrong one (see below). Moreover, by the loss of relevance of the dimensional analysis and the split software/hardware, this approach misses the *proper dimensionality* and the *radical materiality* of biological entities. These are made of that specific matter, the bases of DNA, the molecular components of membranes and … nothing else, in a space that we strictly understand in three-dimensions. There is no way to transfer the biological "information" in DNA on Lego, like in the toy Turing Machine constructed in homage to Turing in Manchester in 2012, and have it work for ontogenesis. Synthetic biology extracts and re-combines fragments of DNA and places them in cellular membranes with their proper physico-chemical and dimensional structure. The dualistic perspectives, software vs. hardware, or soul vs. body, a fantastic invention for the purposes of computing with machines, or a strong religious commitment, respectively, constitute a major distortion of knowledge when imported into the natural sciences. They set on fuzzy grounds, for instance, the analysis of the causes of cancer (see below).

---

8   Wolfram and his followers claim that the Universe may be seen as a (big) Turing Machine (Wolfram, 2013). From this perspective, an apple would fall because it is programmed to fall, like a falling apple on a computer screen. The modern physics of symmetries (see the previous note on Symmetries) is not much affected by these claims. But, for lack of a theory of organisms, the myth that an embryo develops since it is programmed to do so has been more successful than the unneeded computational explanation of falling bodies.
9   Through "mean field theory", in physics, we know that more than three space dimensions force a mean field and forbid singularities, such as barriers, membranes ... a difficult world for organisms. In two dimensions, it is hard to have ducts and their crossing . We seem to be fit just for three space dimensions, no more, no less. Encodings miss or bypass this fundamental aspect of topological/geometric structures: indeed, everything "geometric" or spatial is *sensitive to coding* and to *dimensions*.

## 2. Strong Consequences of Weak Hypotheses

Once focusing on term re-writing as the programming structure of selfish genes, (physical) causality gets out and "instructions" or "recipes" (Maynard-Smith, 1999) guide the analyses of biological phylogenetic and ontogenetic dynamics. So, François Jacob explicitly identified genes with alphabetic writing[10], while W. Gilbert (1992) claimed that, once fully decoded the human DNA, we would have been able to encode it in a CD-rom and say: "Here is a human being, this is me". In the same vein, Francis Collins, director of the National Human Genome Institute, publicly asserted in 2000: "We have grasped the traces of our own instruction manual, previously known to God alone."

The informational approach to biology transforms into ontologies the images of programming on discrete data types as drawn from common sense understanding – with which science is supposed always to break (Bachelard, 1940), like when we moved away from the common sense idea of sunrising on a immobile Earth. As a matter of fact, the reference to "information" and "programming" is not scientific, as it does not use the fundamental properties, nor the fundamental invariants of these robust scientific disciplines, such as the fantastic split software/hardware and the one-dimensional encoding or other mathematical invariants proper to information and programming. Instead, it uses a vague, common sense "transfer" and "weak" meanings[11]. Nor it is metaphorical, as metaphors are rich of meaning transfer; that is, they add knowledge by referring to (other) meaningful contexts. The crude, naive dualism and immateriality of these vague references is sufficient though to erase the singularity and historicity of the living, which can be always surmised as *this living thing here*, in *this three dimensional space*, with *this* body and *this* history.

This specificity of organisms is hard to be described by the ideal invariance of mathematics, by the a-historic and generic nature of its objects; this is implied by the absence of the invention of new mathematical concepts and structures inspired by biology, when compared with the fantastic role of physics in producing new mathematics. And in no way it can be reduced to the uni-dimensional and immaterial invariance of computer software and its digital coding.

In summary, is information used in Shannon-Brillouin sense? Does information refer to Turing-Kolmogorof Algorithmic information theory? Is it always to be viewed as software on discrete data types? In spite of the lack of a scientific specification of what information means exactly, the informational approach was justified by and/or implied several important consequences. First, the molecular structures became the obvious discrete data types and codes for programs and the ultimate information storage of organisms and all biological dynamics. Then the functional specificity of nucleic acids was supposed to be entirely due to individual sequences of its bases, as complete codes for the sequences of the amino acids of proteins. Moreover, exact macromolecular specificity, e.g. the *key-lock* paradigm, was *derived* from the analysis of how to elaborate and transmit information: "Necessarily stereospecific molecular interactions explain the structure of the code ... a boolean algebra, like in computers" (Monod, 1970). Stereospecificity allows the "oriented transmission of information", as assumed by Crick's 1958 Central Dogma of molecular biology, (Monod, 1970).

Here are the shared views that still now follow from the information theoretic frame, as summarized

---

10 "La surprise, c'est que la spécificité génétique soit écrite, non avec des idéogrammes comme en chinois, mais avec un alphabet" F. Jacob, Leçon inaugurale, Coll. France, 7 mai 1965.
11 In a rare attempt to turn these "metaphors" into precise notions, Maynard-Smith (1999), in an extensively quoted paper, explicitly mentions Turing-Kolmogorof (Elaboration or Algorithmic Information Theory) and Shannon-Brillouin (Transmission or Communication of Information), but confuses these approaches in their dual relation to complexity and entropy, see (Longo et al., 2012; Perret, Longo, 2016) for a critique.

in the Stanford's "Biological Information" chapter[12]:

  i. The description of whole-organism phenotypic traits (including complex behavioral traits) as specified or coded by information contained in the genes,
  ii. The treatment of many causal processes within cells, and perhaps of the whole-organism developmental sequence, in terms of the execution of a *program* stored in the genes,
  iii. Treating the transmission of genes (and sometimes other inherited structures) as a flow of information from the parental generation to the offspring generation.

As unambiguously synthesized in (Griffiths, 2001, pp. 395–96) "Genes are instructions—they provide information—whilst other causal factors are merely material…. A gay gene is an instruction to be gay even when [because of other factors] the person is straight."

Under this computational perspective, the informational cascade from DNA to phenotypes is centered on molecular exact ("stereospecific") interactions, that would become the only way ("it is necessary") to transmit and elaborate information, as in a re-writing system. The boolean, key-lock model refers to a formal chemistry that may be analyzed in terms of computational re-writing processes: these transform sequences of letters into sequences of letters, following the instructions, in a deterministic and predictable way, plus some unavoidable noise, (Monod, 1970)[13]. Typically, this excludes physical stochasticity from being an essential component of gene expression, and in general, stochastic and low-affinity macromolecular interactions, whose probabilities depend on the context. This exclusion is contrary to evidence on these chemical phenomena, which dates back to the late '50s (see Kupiec, 1983; Elowitz et al., 2002; Paldi, 2003; Raj et al., 2006 and 2008; Fromion et al., 2013; Marinov et al., 2014). As a matter of fact, since long, chemistry deals with macromolecular interactions in stochastic terms (Gillepsie, 1977). Macromolecules have large enthalpic quasi-chaotic oscillations, are "very sticky" (low affinities are relevant): they are thus treated in probabilistic terms - see the references above and (Creager, Gaudillière, 1996 ; Kupiec, 1996) for more on the origin of this debate. In short, the chemical analyses are based on the *global stochastic* behaviour of populations of macromolecules and not the individual behaviour of each of these molecules, which remains submitted to the perturbing influence of thermal agitation and other "random" dynamics such as affinities with low probabilities. On these grounds, a recent research track, derived from chemistry, radically departs from the "information-programming" approach, where each gene would act like a Laplacian demon "instructing" molecules, individually. The aim is to find a right coarse graining description for understanding the "regulated" stochasticity of macromolecular interactions in a cell, including gene expression. This is based on a mesoscopic level of analysis, of networks in particular, whose dynamics cannot be derived by the knowledge of the constituting elements and which display "canalized" stochastic behaviors, (Giuliani, 2010). This perspective extends to the macromolecular level Boltzmann's approach by microstates in statistical physics, (Kuznetsov, 2002).

The informational language, instead, constructs an autonomous conceptual universe independent from the underlying physical processes and their causal structure: thus, causes are replaced by information flows, signals, control, … programs, whose necessary physical support is the assumed exact complementarity of keys and locks, hands and gloves. Of course, some randomness cannot be excluded. Then, in view of the predictable determinisms of boolean re-writing systems, it is just "noise" affecting, in particular, evolution. Also in this respect then, the information theoretic terminology is not neutral. In particular, it sets a bias on the understanding of biological variability, adaptivity, diversity: they are (or are derived from) noise, to be eliminated or, at best, averaged out, thus treated by "central limit" theorems like in the more recent approaches by Noise Biology, see (Bravi, Longo, 2015) for a critique. In this programming frame, some have been looking for biological novelty even in Gödel's formal combinatorics of signs (see the next footnote). This is not an idea conceived by extremists, but by a few coherent molecular biologists who consistently search for arithmetic recursion and *logical negation* in DNA coding (they are needed to encode Gödel's theorem); it is a tentative, more rigorous "DNA as a program" approach that, at least, goes beyond the usual vague, common sense use of "information" and "programming" in biology[14].

---

12  Philosophy of Biology:  http://plato.stanford.edu/archives/fall2008/entries/information-biological/

13 "Biological specificity ... is entirely ... in complementary combining regions on the interacting molecules" (Pauling, 1987). "The orderly patterns of metabolic and developmental reactions giving rise to the unique characteristics of the individual and of its species ... the shapes of individual molecules allow them to selectively recognize and bind to one another. The main principle which guides this recognition is termed complementarity. Just as a hand fits perfectly into a glove, molecules which are complementary have mirror- image shapes that allow them to selectively bind to each other" (McGraw-Hill Dictionary of Scientific & Technical Terms, 6E, 2003).

Thus, as for biological randomness (as unpredictability or production of novelty), information-oriented frames do not need to refer to the complex blend of physical, classical and quantum randomness in a cell: either it is noise in information-elaboration channels or it is … "pseudo-Gödelian". Even the Brownian motion is seen as disturbing noise in exact, Turing-machine like, genetic expression. Brownian motion instead, jointly to the enthalpic oscillations of macromolecules, dominates the physical dynamics and the energetic landscape and has a constructive role in both prokaryote and eukaryote cells (see the references above to stochastic gene expression, an approach harshly marginalized for decades by the informational mainstream; (Richard et al., 2016) presents further experimental evidence). The stochastic approach highlights a fundamental causal component of bio-chemical interactions in continuous dynamics, while "stochastic" coding, information transfer and programming would make little sense. Moreover, different forms of randomness, at all levels of organization, may causally contribute to phenotypic changes and to biological stability by adaptivity and diversity, see (Bravi, Longo, 2015) for references and (Buiatti, Longo, 2013) for the further notion of "bio-resonance".

The informational perspectives do not refer either to additional physical phenomena in cells, such as the possibly very relevant role of the "super-coherence" of water. This is a Quantum

---

14 **On Gödelitis.** In an attempt to bypass Monod's mechanistic-formal approach, enriched by some noise, Danchin (2003) and (2009) tried to bring Gödel's theorem into the picture. Beyond formalism, Gödel's incompleteness would prove the unpredictable "creativity" of biology within the programming approach. A remarkable attempt for a leading biologist, as these issues in Logic are far from common sense (see (Smorinski, 1978), (Longo, 2010), (Longo, 2011a) for technical references). Indeed, (Rogers, 1967), a classic in Computability Theory, calls "creative" the set of (encoded) theorems of arithmetic, i.e. the formal-mechanical consequences of the axioms. By Gödel's first theorem, this set is not computable (and, to the biologist, the word may recall Bergson's Creative Evolution). Yet, this set is *semi-computable*, meaning that it may be effectively generated and, as such, is far from "unpredictable", since an algorithm produces all and exactly all its infinite elements. Moreover, the recursive generation of Gödel's undecidable formula is effective as well: it is an incredibly smart recursive and diagonal construction (it uses logico-formal negation), which allows to *generate* a formula not derivable from a specific set of axioms. This procedure may be indefinitely and effectively iterated. In short, Gödelian undecidability, constructed by an encoding of the metatheory into the formal theory, does *not finitely generate* unpredictable information as the diagonal formula may be effectively produced, even though it is not derivable from the information in the axioms. Thus, formal derivability is *semi*-computability or *semi*-decidability, i.e. the "information" in the axioms does not allow to decide all formulae, for example the diagonal one. Yet, it allows to effectively construct it and to show that … it is not decidable. That is, the construction of the sentence that escapes the given axioms is also effective (semi-computable). Theoretical unpredictability, the least property one expects for "creativity" in nature, is at least (algorithmic) randomness (Calude, 2002) and this is far from semi-computability. It is the "opposite" (e.g. a random set of numbers and its complement cannot contain *any infinite semi-computable* subset) and it may be soundly compared to unpredictability in physics (e.g. it mathematically relates to classical ergodicity and to Quantum randomness, asymptotically (see Calude, Longo, 2015, for a survey)). If creativity, thus unpredictability, were just Gödelian semi-computability and its recursively produced undecidable formula, I would love to have a program generating a semi-computable subset of an unpredictable process, such as future phenotypes or, better, future lottery drawings: this would make me immensely rich. The merit of Danchin's remarks, though, is that they are based on precise notions, thus they may be proved to be wrong.

Electrodynamic effect in highly partitioned structures, such as in an organism made of $10^{13}$ cells, which accelerates the Brownian motion of non-water molecules at constant temperature and enhances the rate of biochemical activity, (Del Giudice, 1983, 1986; Arani, 1995). Both in the information/programming approach to biology and in the Theory of Programming, a robust science on its own, the underlying hardware has no interest for the program analyst, provided that it works correctly, in spite of some noise. In Computer Science, the needs of Programming set the standard of "correct" working for the physical, material structure, which followed by more than 10 years Turing's mathematical distinction between software and hardware. It is the engineers' job to have the hardware work according to the programmer's needs and, thus, realize an interface appearing as a (Turing-vonNeuman) discrete state architecture, with whatever means they have. And they can have it work correctly, which is just fantastic: in modern computers, we implemented the strongest form of Cartesian soul/body split, by radically subordinating matter (hardware) to an independent spirit (software). Similarly, the material cell must follow the genetic instructions; it is an Avatar (see the footnote above on Avatars). Yet, the genome may escape from them and generate novelty internally, independently from physics, from the organism or from the ecosystem, by just implementing Gödel's self-referential encoding of the formal metatheory, a set of instructions on how to diagonalize on alpha-numeric signs (see the footnote on Gödelitis).

Following this extreme Cartesian dualism, organismal biology has thus been reduced to purely formal laws of a symbolic chemistry, a virtual interface handled in terms of information and programming theory, with a reference to physics in occasional reductionist claims[15]. But, if the causal structure of this presumed formal bio-chemistry of information in macromolecules is absent, doubtful or incomplete, which laws of physics are actually refereed to, in reductionists perspectives in biology?

Physics, from Galileo to Quanta, has never ceased to construct and modify its laws by confronting unprecedented phenomena or by novel insights into known phenomena or just by … changing the scale of observation. There is *no reduction within physics*, as it proceeds by "unification", from Newton and Boltzmann to current issues in between Quantum and Classical/Relativistic physics or hydrodynamics, see (Chibbaro et al., 2015) and (Longo, 2016) for a review. For example, hydrodynamics, as a science of incompressible fluids in continua, is not understood in terms of quanta; physicists try instead to invent a new theoretical frame that could unify these theories (note that there is a lot of water in an organism … thus, which "physics" are reductionists in biology referring to?). Moreover, classical and quantum random phenomena, which are far from being unified, are both present and interact in cells and may have phenotypic effects (Buiatti, Longo, 2013). In physics, all existing *unifications* were based on very strong theoretical hypotheses, grounded on revolutionary ideas. For instance, Newton equations and infinitesimal calculus, which unified Galileo's falling stones and celestial bodies; Boltzmann asymptotic construction of Statistical Physics, which unified particles' dynamics and Thermodynamics on the grounds of the ergodic hypothesis, an incredibly strong and precise statement; String Theory or Non Commutative Geometry, as for today's attempts to unify quantum and relativistic fields by incredibly strong, revolutionary assumptions and concepts, surely not derived from common sense. And none of these is a "reduction" to a "lower" level. Moreover, as it is in the two last cases above, unification, in science, should always be provisional and "local," not dogmatic and a priori reductionist, but critically constructed.

The deduction of strong consequences from weak, fuzzy, a-scientific or "common sense" hypotheses, such as the "information" or "programming" assumptions in biology, is unacceptable as a scientific praxis. Note finally that, the pre-scientific reference is only made to a Theory of Information on discrete data types, elaborated, transmitted and encoded by programs, written as alpha-numeric instructions. No reference is ever made, that I know, to the well-established discipline of Geometry of Information, where symmetry changes in possibly continuous symmetry groups propose a radically different conceptual frame (Barbaresco, Djafari, 2015).

---

15 « Life can be explained on the basis of the existing laws of Physics » (Perutz, 1987)

In addition to the few listed above, we will see some specific strong consequences of the weak hypotheses transferred from common sense notions of "information" to biology and how these affect current cancer research. This domain has been for too long dominated by the myth of the computer program, centralized in the DNA (the Central Dogma of Molecular Biology): the focus on information-program-signal, as drivers of development, supports the idea that embryogenesis as well as both normal and pathological development should be always or first studied as a DNA centered, programming issue. In this context, cancer has been consistently analyzed as the result of DNA *de-programming* either inherited or *provoked by a* carcinogen disrupting the DNA *encoded instructions* (the mutagenic effect of a carcinogen). Let's briefly summarize some steps of this still prevailing view of the etiology of this life-threatening disease, a view that recently received further support from the software industry, in spite of massive negative evidence.

## 3. An announced debacle

> "… you cannot prove a vague theory wrong"
> Richard Feynman (1964)

We will follow the story of a wrong path as courageously acknowledged by one of the founding fathers and major actor of the dominating theory, in biology of cancer, R. A. Weinberg, in his 2014's paper (see references). The so called Somatic Mutation Theory (SMT) postulates that cancer originates as a one-cell disease, thus it would then be clonal, and would be due to one or more mutations (driver mutations). All of these features would be either a consequence of a chemical reaching such a cell from a carcinogen or else due to hereditary causes (see (Nowel, 1976, Cairns, 1981; Strauss, 1981) for classical surveys of this century-old theory, originally proposed, in a different language, by (Boveri, 1914)).

Since 1971, generously funded projects have heralded the final victory against cancer thanks to genetic therapies able to "reprogram" the "deprogrammed DNA", within a few years. In particular, this approach was at the core of President Nixon's War on Cancer (see below for more quotations on this). The common sense notion of "program" was indeed understandable also by Nixon; a major advantage of using an everyday, a-scientific language, as this facilitates the message to be understood by everybody. Moreover, programs can be debugged, thus the promise of genetic therapies as DNA debugging (see below). In spite of providing neither plausible explanations of the carcinogenic process nor therapeutic concrete solutions, since 1971, a major technological achievement, by the year 2000, , i.e., the complete decoding of human genome, was used to offer further tools to solve the cancer puzzle and generate, once again, genetic therapies for cancer. These had to be expected at latest within 10 or 15 years, while sound diagnosis and prognosis were promised much sooner on the grounds of newly uncovered fundamental "hallmarks" of cancer. Genetic analysis of cancer cells should have provided diagnosis of malignant vs benign forms of this disease, primary vs. metastatic cancers etc. These optimistic papers are too many to be listed; it may be enough to quote (Collins,1999), the head of the Genome Project, (Hanahan, Weinberg, 2000) (over 20,000 quotations in a few years), (van Eschenbach, 2003), all major personalities in the field. In (van Eschenbach, 2003), cancer is viewed both as "a genetic disease and a cell signaling failure. Genes that control orderly replication become damaged"; on the grounds of this causal analysis, the paper promises, by 2015, genetic therapies for "eliminating suffering and death due to cancer". Incidentally, this claim was supported by the American Association for Cancer Research in 2005.

Thanks to the full knowledge of DNA sequences in normal and cancer cells, these proposed upcoming therapies are supposed to be based "on scientific laws as robust as those of chemistry and physics" (Hanahan, Weinberg, 2000). The proximity of metaphors of "programming" to common sense, as always, facilitated these promises among funding agencies and among the general public. The enormous financial efforts and the ruthless exclusion of alternative hypotheses have both been motivated for decades by the general idea that any phenotype presupposes its complete determination by the genes. However, a half-century of genetic research has produced no plausible gene-based cancer therapy, see (Baker, 2014; Huang, 2014), two elegant syntheses and highly recommendable reading to the non-biologist (but so worrying!)). As Weinberg (2014) himself acknowledges "We were, after all, reductionists, who would parse cancer cells down to their smallest molecular details and develop useful, universally applicable lessons about the mechanisms of cancer development … Half a century of cancer research had generated an enormous body of observations about the behavior of the disease, but there were essentially no insights into how the disease begins and progresses to its life-threatening conclusions". So, Weinberg observes that "a particularly jaundiced cancer researcher" commented to him that "one should never, ever confuse cancer research with science!".

How could DNA be de-programmed according to the early research projects? At the beginning of the 1971 War on Cancer, retroviruses were considered as DNA de-programming agents. "Few seemed deterred by the well-established observation that most types of human cancer did not represent communicable diseases" (Weinberg, 2014). Ramazzini, anatomist and physician in Bologna had already made this observation in early XVIIIth century[16]. Weinberg continues his auto-critique (pp. 267-9) by summarizing further spurious key steps in the SMT approach to cancer. Since 1973 the search focused on "chemical species correlated directly with mutagenic activity" . He then recalls the progressive move, between 1982 and 1999: from "just one mutation" to "a specific sequence of mutations".  "Only later was it clear that most human carcinogens are actually not mutagenic ... but fortunately I and others were not derailed by discrepant facts" (sic). This is a crucial remark. As a matter of fact, there is increasing evidence that many (most?) carcinogens interfere on tissue organization, not by sending (chemical) signals that de-program DNA. For example, Maltoni (1980) observed the disruptive role of asbestos micro-filaments on the tissue matrix, on cell connections and membranes, but could not point to any direct mutagenic effect. This observation was in contrast with the claims of the dominating SMT and, hence, received little consideration. As a matter of fact, when asbestos is made into powder, it ceases to generate cancer "fiber dimension is one of the important determinant factors of asbestos carcinogenicity" (Huang et al. 2011). Also, by subcutaneously inserting diverse inert objects (plastics, metals, etc) it has been shown that their carcinogenic effects depended not on their chemical make up but on their peculiar physical structure (e.g. the carcinogenic effect may depend on the presence and size of micropores in plastic membranes, a fact known since (Karp et al., 1973)). Of course, mutations will follow as consequences (passenger mutations) not causes (driver mutations) of cancer, see below.

Other commentators of note have expressed their views on carcinogenesis for the record. In a very interesting interview, Venter (2010), whose team first decoded the human genome in 2000, acknowledged that "'We Have Learned Nothing from the Genome". Wrong expectations were due to "the ill-founded belief that those who know the DNA sequence also know every aspect of life … That is nonsense". However, cancer biologists did learn something from the Genome Project. The extensive decoding of the DNA of cells in cancerous tissues showed that, in the same tissue, cells may have very different mutations and chromosomal changes:  "Genome sequencing also came of age and documented myriad mutations afflicting individual cancer cell genomes" (Weinberg, 2014). More precisely, "63 to 69% of all somatic mutations [are] not detectable across every tumor region … Gene-expression signatures of good and poor prognosis were detected in different regions of the same tumor" (Gerlinger et al., 2012), see also (Kato el al., 2016). No much help came from genomics in the analysis of metastasis, as acknowledged also by proponents of SMT: "Despite intensive effort, however, consistent genetic alterations that distinguish cancers that metastasize from cancers that have not yet metastasized remain to be identified … The idea that growth at metastatic sites is not dependent on additional genetic alterations is also supported by recent results showing that even normal cells, when placed in suitable environments such as lymph nodes, can grow into organoids, complete with a functioning vasculature" (Vogelstein et al., 2013). In the interpretation hinted in the next section, normal cells in a context that cannot control and canalize their "normal" reproduction with variation may yield a "pathological" situation. Moreover, no driver mutations specific to metastasis have yet to be documented (Zhang et al., 2013: Alshaya et al., 2014; Versteg, 2015).

---

16  The papillomavirus and the HBV-HCV viruses (hepatitis), associated with cancer, are not retroviruses.

Finally, it is remarkable that cells in healthy tissues may have the genetic hallmarks of cancer: "aged sun-exposed skin is a patchwork of thousands of evolving clones with over a quarter of cells carrying cancer-causing mutations while maintaining the physiological functions of epidermis" (Martincorena et al., 2015). Equally noteworthy is that cell aneuploidy and polyploidy, that used to be considered as another chromosomal signature of cancer are present in 50% or more normal liver cells and are considered to be beneficial by assuring resilience to toxic shocks and for liver regeneration (Duncan, 2013).

Following the quotations referred above, a few relevant facts have become clear from the massive DNA decoding of cells in cancer tissues. They are:

1 - Gene-expression signatures for benign and malignant cancer may coexist in the same tumor.
2 - Genetic analyses do not allow to discriminate between a tumor that (has or) will metastasize(d) from another that (has or) will not.
3 – DNA sequencing does not help in distinguishing a primary from a metastatic cancer.

Note that 90% of lethal cancers are metastatic (Sporn, 1999; Cook, 2011). This stresses the relevance of the last two points. Of course, the etiology of cancer remains open, that is, the origin of primary cancers. Yet, proponents of SMT acknowledge that 99.9% of mutations found in cells of all cancer tissues are passenger not driver mutations of cancer, see (Vogelstein et al., 2013) and the next section[17]. So, in a more than vast majority of cases, many seem to acknowledge that the "primary and immobile motor" of ontogenesis (and thus of cancer as of any phenotype), DNA as a program, becomes a passive recipient of orders (the passenger mutations). Of course, the messy situation of cells' chromosomes in a cancer (not just mutations, but massive polyploidy, aneuploidy, etc) negatively retro-acts on tissues' healthy dynamics: their deregulating effects may even further disrupt the cells' dialogue, hormonal control of reproduction etc. see the next section and (Sonnenschein, Soto, 1999, 2011; Baker, 2014, 2015; Huang, 2014) for surveys.

In view of the remarkable empirical knowledge that DNA decoding has provided, are we approaching the end of a (de-)programming DNA centered view of cancer and of ontogenesis in general? Hopefully, empirical negative results in the natural sciences should have the same role as "negative results" in mathematics or mathematical physics: in principle, they should modify scientific thinking, scientists may become more open to or invent new theories, new scientific paradigms (Longo, 2010). To the contrary, the genocentric informational views cannot be falsified by experience, because they are not scientific: those views are based on common sense notions of information and program and on the "homunculus" ancient believe, modernized by encoding it in chromosomes. Thus, the massive presence of mutations and chromosomal alterations in cancer tissues continues to be perceived as *the cause* of the disease, since according to that theory any phenotype must have an antecedent in the genotype and the genotype is supposed to completely control the organism-avatar. We know from human history that when common sense and myths combine, they are invincible or any change requires a true revolution.

---

17 In reference to the percentages mentioned in the last few lines, it may be fair to claim that the vast majority of research funding (90% ?) in biology of cancer, in the last few decades, has been allocated to geno-centric approaches and that most publications in biology of cancer (90% ?) is still now devoted to those analyses. These two aspects of research trends are also the result of the amplifying effect of bibliometrics, that reinforces main stream, fashionable areas (Longo, 2014), and thus enhances positive retro-actions between funding and publications.

Following the trend, Microsoft proposes to help in solving the cancer puzzle by its technical (or commercial?) skill in software production: Microsoft's "computing cancer project" (2016) claims that one has to understand how the cell's programs work, then "If you can figure out how to build these programs, and then you can debug them, it's a solved problem". Their motto is "Our approach to solving cancer: debug the system"[18]. Is this just surplus money that goes to cancer research? Not necessarily, because joint ventures in this enterprise apply for funds to research institutions. And, more importantly, Microsoft's talent for commercials and publicity, which are the actual aim of these announcements in spite of the sufferings they refer to, may confirm *common sense* by reaching politicians and managers who decide about funding; in short, it sets a reference. IBM also offers DNA decoding services for cancer diagnosis and prognosis, in spite of the evidence mentioned above. And Big Data enter massively in the game. In view of the very heterogeneous and unexpected genetic situation of cancer cells, of the "myriad mutations afflicting individual cancer cell genomes" thus of "cancer's infinite complexity" (Weinberg, 2014), and of the failure to turn cancer biology into a science, many researchers follow (Anderson, 2008) philosophy. Namely, collect all "-omics" available data (genomics, proteomics, metabolomics …), then "... throw the numbers into the biggest computing clusters the world has ever seen and let statistical algorithms find patterns where science cannot ... Correlation supersedes causation, and science can advance even without coherent models, unified theories … No semantic or causal analysis is required". Of course, the larger is the database, the best for prediction and action without understanding.

We are coming full circle back to the more than 100 years old remarks by Riemann, Jeans and others quoted above: if you have only discrete manifolds, give up causality. Thus, consistently claim the purest Data Miners, just look for correlations without explanations – science is no more needed. Note that these provably wrong claims against theorizing may neglect measurement as well: classical, Riemannian and quantum challenges as for physical measurement are forgotten – a digital database is exact, the metrics is intrinsic. The pre-given discrete structure of the databases may thus help to forget *how* these data have been collected in complex biological organisms. The (often implicit) a priori's in the choice of observables and of their metrics do not need to be discussed, as this would be "theorizing" (indeed, Data are "Compressed Theories", as their "collecting" supposes a theoretical perspective, (Longo, 2016)).

Now, as formally shown in (Calude, Longo, 2016), sufficiently large sets of numbers, even when produced by a random process, necessarily contain correlations. More precisely, a nice and not-obvious combinatorial theory of numbers, i.e. Ramsey Theory, proves the following:

> *(Informal)* Set the criteria for a correlation in a database: its n-arity (you want to correlate n variables), the length p of the correlation (you want it to be long enough, e.g. n data must correlate every minute, for a year, say), the number c of parts you divide your database (you give the same "color", say, out of c colors, to numbers that you consider correlated: they are close or happen simultaneously or whatever). Then, one can compute a number, d say, such that f*or* any set A with d elements or more and f*or* any partition of the n-uples in A in c colors, there exists a subset B of A that contains p elements and is monochromatic, i.e. it is entirely contained in one partition.

The number d above is truly "huge", but isn't the larger the best? Then the data miner may happily exclaim: "we have got a correlation!", even when … the data set A has been produced by a random generator. Indeed, in immense numeric databases one has a deluge of spurious correlations, in a very strong sense, as the set A above is arbitrary. Thus, A may have been obtained by … throwing dices, flipping coins, quantum measurements … arbitrary choices of observables and measurements. It is hard to predict and act on these grounds. Moreover, when you are dealing with very large sets of numbers, most of them are "random", in a precise sense (Calude, Longo, 2016). It may be wiser, then, to try some scientific theorizing.

---

18 As a former user, now a Linux fan, I think that Microsoft should better and first debug its own software, see (Di Cosmo, 1998).

## 4. Towards TOFT

Following a different research path, an approach proposed by cancer biologists Sonnenschein and Soto (TOFT, Tissue Organization Field Theory, see the references by these authors) is based on Darwinian principles that we further extended to a tentative theory of organisms (see the next section). The TOFT approach to cancer refers to early intuitions by C. Waddington, J. Needham and a few others (1930s), later forgotten by the subsequent genocentric perspective (see (Sonnenschein, Soto, 2011) for references). The novelty and the suitable "paradigm instability" brought in by TOFT vs SMT is analyzed in (Baker, 2014; 2015) and (Smythies, 2015).

TOFT key principle is that all cells, including somatic cells, tend a priori to reproduce: in (Sonnenschein, Soto, 1999) terminology, *cell proliferation* and *motility* is their "default state". We extended this default state to the idea that all organisms as well as the cells in multicellular organisms, tend to reproduce *with variations* and to move, as more closely spelled out in the next section. This is an extension to cells within an organism of Darwin's principle of heredity in evolution as "descent with modification", which occupies three out of the first six chapters of the Origin of Species (see Longo et al, 2015; Montévil et al., 2016) . This revolutionary principle is essential to Darwin's second principle, *selection*. It is a "limit-state" analogous to Galilean inertia, but specific to life forms. Note that inertial movement is a *limit* principle, as it is always constrained and modified by gravitation and frictions. Analogously, somatic cells, and also organisms in an ecosystem, are constrained/controlled by the organism or the environment in their free reproduction and movement. As Darwin observes, an unconstrained organism would quickly cover the entire Earth, by reproduction. Galileo's inertia, Darwin's principles and the default state of reproduction with variation and motility are all derived from observation and posed as principles of intelligibility at the core of their theoretical approach. By positing inertia, asymptotically (no physical body moves like a point on an Euclidian straight line at constant speed), Galileo could analyze what affects it, gravitation and frictions. On the grounds of his first principle, Darwin could propose selection as acting on organisms. TOFT central idea then is to analyze what controls cell reproduction with variation and motility in an organism (see (Longo et al., 2015; Soto et al., 2016) for more on Darwin and the conceptual analogy with Galileo's principle of inertia). Under this perspective, cancer is a tissue-based, organismal problem, akin to the process of morphogenesis during development: "cancer is development gone awry" (Sonnenschein, Soto, 2011).

In summary, within an organism, when effective control by intercellular exchanges, tissues matrix, hormones, etc. is disrupted by a carcinogen, cells reproduce and change at a speed that may even reach that of embryogenesis. This in turn modifies the micro-environment, it actually complexifies it in a precise histological sense, while reducing tissue (organ) functionality, an hallmark of cancer as we observe in (Longo et al., 2015). Note that, in contrast to the claim by the SMT that "once a cancer cell, always a cancer cell", cells from a mammary carcinoma (an epithelial cancer), when placed into a normal mammary stroma (the normal micro-environment of the mammary epithelium) revert to normalcy (Maffini et al. 2005). The idea is that cancer does not depend on a "triggering signal" at the molecular level, which would deprogram the DNA of an *a priori* quiescent cell by inducing a driver mutation. Instead, cancer can be considered as the failure of the regulatory relations between and of cells in a tissue and of the tissue in an organism. Passenger mutations massively follow (also for SMT supporters, they are 99.9% of mutations in cell in cancerous tissues, see above), as mutations are the main way to generate variation at the cellular level. These hypotheses, and their therapeutic consequences redirect the attention of researchers toward prevention and modifications of environmental conditions; in particular, towards reconstructing the cells' micro-environment, (Cook et al., 2011; Bizzarri, Cucina, 2014). In the latter case, like in the recombination experiments in (Maffini et al, 2005), cells inside a cancer can be normalized. The reader should consult (Baker, 2014; Smithies, 2015; Pisco, Huang, 2015) for surveys: "Thinking in terms of TOFT can spur new lines of research"(Baker, 2015). Also, many if not most cancer "conundra" are made understandable along these new lines of thought, (Kato et la., 2016).

**5. From TOFT to Working Hypothesis in Biology of Organisms**

From our general attempt, we conclude that the rich knowledge construction proper to physics is not lost. In two books (Bailly, Longo, 2011; Longo, Montevil, 2014) and several papers, we have tried to articulate certain physical and mathematical theories with phenomena that are specific to life and worked on some specific "perspectives" on organisms (rhythms, biological time, criticality ...). We then joined the efforts of our colleagues, Sonnenschein and Soto towards the proposal of a "theory of organisms", introduced in (Longo et al 2015) and summarized in the volume (Soto, Longo, 2016), in particular in both (Soto et al, 2016) papers. In our approach, DNA is a fundamental, internal "constraint" to cellular and biological activity, where we used constraints in the sense described in (Montévil, Mossio, 2015; Mossio et al., 2016). That is, DNA is a physico-chemical trace of an entire history (Longo, 2017), continually used by the cell dynamics, and thus constraining it to certain proteomics, according to the context (beginning by the boundaries it sets to the proteome's Brownian motion, possibly enhanced by quantum effects, see above). In this regard, (Montévil et al, 2016b) modeled mammary gland morphogenesis by the dynamics of constraints that, generated by the cell agency, organize the surrounding matrix, which in turn, constrain the proliferation and motility of the cells.

In this frame, it is appropriate to go back to Darwin, whose greatness is to have formulated autonomous theoretical principles of intelligibility of phylogenesis, on the principial model, but not the techniques, of the major creators in mathematical physics. As mentioned above, the two Darwinian principles of evolutionary heredity are *descent with modifications* and *selection*. The current challenge is to articulate these principles with the analysis of the organism, in the long term attempt to unify ontogenesis and phylogenesis. The role of strong, explicit principles in mathematics and physics is crucial. In (Longo, 2015), from which this section is partly borrowed, Euclid's "line with no thickness" (a definitional principle made explicit in definition β, book I) and Galileo's principle of inertia are extensively discussed. They are limits, that is the infinite limit of decreasing thickness and a limit movement, respectively, as well as founding principles for knowledge construction, far away from *common sense*. Our quest for principles in biology follows these examples, while acknowledging that the principles specific to physics—grounded on invariance, conservation properties as symmetries, and optimal trajectories—are insufficient for the proper observables of living beings, organisms and phenotypes. Living systems are, instead, in a permanent state of *critical transition*: their symmetries are continually breaking up and being reconstituted, at least at each cell reproduction (Bailly, Longo, 2011; Longo, Montévil, 2014; Longo, Soto, 2016). In our perspective, Darwinian principle of *reproduction-with-variation* may be seen as a principle of *non-conservation*, opposed to and symmetric with the principles of conservation and invariance in mathematics and physics, but at the level of the appropriate biological observables, that is, organisms. The adequate theorization of the biological field therefore demands extensions and sums of various physical theories— such as the ones due to the coexistence of random classical and quantum phenomena in the cell (Buiatti, Longo, 2013), far from equilibrium dynamics (Nicolis, Prigogine, 1977), extended criticality (Bailly, Longo, 2011; Longo, Montévil, 2014). These operations rely on physical theories and extend their methods, while remaining irreducible to their mathematical techniques. They propose proper biological principles as well as "points of view," and "perspectives" on the organism, whose unity furnishes the guiding thread through these different theoretical aspects. The intelligibility of the biological field is only possible through intersections and partial integrations that aim to construct objects-of-knowledge in dialectical relation with the constraints of experience. In biology, experiences plays a singular role, beginning with the difference *in vitro* vs. *in vivo*, unknown to physics, and the peculiar role of historical knowledge and, thus, of diachronic measurement in theory building (Longo, 2017). Unity with physical theories (classical, quantum?) may be a long term goal, surely not a reduction as hinted at the end of sect. 2.

Thanks to mathematization, theorizing in physics extracts generic objects and properties, out of intentional observations and measurement, as conceptual and mathematical invariants. Their objectivity as invariance depends entirely on the theoretical framework**.** In biology, instead, objects are always historic singularities, which are grasped by conceptual models that are qualitative, provisional, and over-determined by history and cultural perspectives. The centrality of each singular organism, with its own historicity, implies the primacy of variation and symmetries' breaking that overthrow the current mathematical primacy of invariance—a primacy with very powerful knowledge effects, but which may prove an obstacle to understanding life, especially when it is disfigured in the genocentric approach to DNA and the myth of the "program", as the informational invariants. For example, the radical materiality of organisms that we mentioned, its historical thickness, and the density of its internal and external relations, rule out any dualism between "software" and "hardware", discussed above. Finally, one of the very conditions of possibility for physical knowledge, the space of phases (the observables and the parameters), is overthrown in biology. In physics, the (phase) space is fixed a priori, a proper one for each physical theory (classical, quantum, hydrodynamics, thermodynamics …), as the Kantian condition of possibility and immanent norm of physical trajectories. In biological processes, by contrast, the phylogenetic trajectories constitute and constantly reorganize the space of possibles (of phases), the ecosystem. The observables (phenotypes and organisms) are the *results* of the processes. The historicity of life is grounded on these changes of observables and parameters along evolution (phenotypes and pertinent parameters change), and on the key role of *rare events*, a peculiarity of historical processes, (Longo, 2017).

If our analysis of living dynamics is pertinent, it poses the problem of how to test the limits of traditional scientific objectivities, of which physics and mathematics represent the paradigms, in the face of the constraints of biological theorization. Overcoming very powerful theoretical practices that are rooted in old, deep and very powerful metaphysical and *theological* ideas, (Longo, 2011b), is a radical challenge, but some attempts are seeing the light of day, ours is one of them.

**Acknowledgments**. Stuart Baker, Alessandro Giuliani, Carlos Sonnenschein and Ana Soto encouraged and commented this paper on cancer related issues. Alastair Abbott made several comments on Quantum causality (see Abbott and Calude's enlightening writings in this Blog on Quantum Computing:
http://www.quantumforquants.org/quantum-computing/limits-of-quantum-computing/  )

**References** (Longo's (co-)authored papers are in http://www.di.ens.fr/users/longo )

Aceto L., G. Longo, B. Victor (Editors), **The Difference between Concurrent and Sequential Computations,** *Special issue*, Mathematical Structures in Computer Science, Cambridge U. P., vol.13, n.4 - 5, 2003.

Alshaya W, Mehta V, Wilson BA, Chafe S, Aronyk KE, Lu JQ. Low-grade ependymoma with late metastasis: autopsy case study and literature review. **Childs Nerv Syst.** 31(9) · May 2015.

Anderson C. "The end of theory: The data deluge makes the scientific method obsolete" *WIRED*. 2008.

Arani R, Bono I, Del Giudice E., Preparata G., QED coherence and the thermodynamics of water, **International Journal Physics** B9, 1813, 1995.

Asperti A., Longo G.. **Categories, Types and Structures**. M.I.T. Press, 1991.

Baccelli F., Mir-Omid Haji-Mirsadeghi, Ali Khezeli, Dynamics on Unimodular Random Graphs, on r arXiv:1608.05940v1 [math.PR], 2016a.

Baccelli F., Mir-Omid Haji-Mirsadeghi, "Point-Map-Probabilities of a Point Process and Mecke's Invariant Measure Equation", on arXiv:1312.0287v3 [math.PR], 2016b.

Bachelard G. **La Philosophie du non**, PUF, 1940.

Bailly, F., Longo, G., **Mathematics and the natural sciences: the physical singularity of life**. London: Imperial College Press, 2011 (original French version, Hermann, 2006).

Baker S., "Recognizing Paradigm Instability in Theories of Carcinogenesis", **British Journal of Medicine & Medical Research,** 4(5): 1149-1163, 2014.

Baker S., "A cancer theory kerfuffle can lead to new lines of research". **J. Natl. Cancer Inst**. 107, dju405, 2015.

Barbaresco F., Mohammad-Djafari A. (Eds.) , **Information, Entropy and Their Geometric Structures**, MDPI, Basel & Beijing , 2015.

Bezem M., J. W. Klop, R.. Roelde Vrijer. **Term Rewriting Systems.** Cambridge: Cambridge U. Press, 2003.

Bizzarri M, Cucina A., "Tumor and the microenvironment: a chance to reframe the paradigm of carcinogenesis". **Biomed Res Intl.** 2014:934038, 2014.

Boveri, T. **Zur Frage der Entstehung maligner Tumoren**. Jena: Gustov Fischer, 1914.

Buiatti M., Longo G. "Randomness and Multi-level Interactions in Biology". In **Theory in Biosciences**, vol. 132, n. 3:139-158, 2013.

Bravi B., Longo G. "The Unconventionality of Nature: Biology, from Noise to Functional Randomness" **Unconventional Computation and Natural Computation,** Springer LNCS 9252, Calude, Dinneen (Eds.), pp 3-34, 2015.

Cairns, J. The origin of human cancers. Nature 289:353–357,1981.

Calude C. **Information and randomness**. Springer-Verlag, Berlin, second edition, 2002.

Calude C., Longo G. "Classical, Quantum and Biological Randomness as Relative Incomputability". *Invited Paper*, special issue of **Natural Computing**, Springer, to appear, 2015.

Calude C., Longo G., "The Deluge of Spurious Correlations in Big Data". *In* **Found. of Science**, 1-18, March, 2016

Chibbaro S., Rondoni l., Vulpiani A. **Reductionism, Emergence and Levels of Reality: The Importance of Being Borderline,** Springer, Berlin, 2015.

Collins F., "Medical And Societal Consequences Of The Human Genome Project", **The New England J. of Medicine,** Jul y 1, 19 9 9.

Cook LM, Hurst DR, Welch DR. "Metastasis suppressors and the tumor micro-environment." **Semin Cancer Biol.** 21(2):113–122, 2011.

Creager A. N. et Gaudillière J.-P., "Meanings in Search of Experiments and *Vice-versa* : the invention of allosteric regulation in Paris and Berkeley, 1889-1968", **Historical Studies in the Phisical and Biological Sciences**, 27, 1-89, 1996.

Crick, F.H.C., "On Protein Synthesis". **Symp. Soc. Exp. Biol**. XII, 139-163, 1956.

Danchin A., **The Delphic Boat. What genomes tell us**, Harvard University Press, 2003.

Danchin A., "Bacteria as computers making computers", **Microbiology Review**, 33: 3–26, 2009.

Del Giudice E., Doglia S., Milani M., Vitiello G.: Spontaneous symmetry breakdown and boson condensation in biology, **Phys. Lett.**, 95A, 508, 1983.

Del Giudice E., Doglia S., Milani M., Vitiello G.: Electromagnetic field and spontaneous symmetry breakdown in biological matter, **Nucl. Phys.,** B275, 185, 1986.

Di Cosmo R., **Le hold-up planétaire**, Calmann-Lévy, Paris, 1998.

Duncan A. "Aneuploidy, polyploidy and ploidy reversal in the liver". **Semin Cell Dev Biol.** Apr; 24(4):347-56, 2013. doi: 10.1016/j.semcdb.2013.01.003. Epub 2013 Jan 16.

Elowitz, M. B.,Levine, A.J., Siggia, E., Swain, P.S.: "Stochastic Gene Expression in a Single Cell". **Science**, 297, 2002.

von Eschenbach A., "NCI Sets Goal Of Eliminating Suffering And Death Due To Cancer By 2015", **J Natl Med Assoc.,** 95:637-639, 2003.

Fox Keller E., **The century of the gene**, Harvard U. P., 2000.

Fromion V., E. Leoncini, and P. Robert. Stochastic gene expression in cells: A point process approach. **SIAM Journal on Applied Mathematics**, 73(1):195–211, 2013.

Gerlinger M. et al (22 authors). "Intratumor Heterogeneity and Branched Evolution Revealed by Multiregion Sequencing", **Engl J Med** 366;10, march 8, 2012.

Gilbert W. "A vision of the Grail" *in* **The Code of Codes: Scientific and Social Issues in the Human Genome Project** (Daniel J. Kevles, Leroy E. eds) Harvard U.P., 1992.

Gillespie D. Exact stochastic simulation of coupled chemical reactions. **J. physical chemistry,** 81(25):2340–61, 1977.

Giuliani A., "Collective motions and specific effectors: a statistical mechanics perspective on biological regulation" **BMC Genomics**, 11(suppl 1):S2, 2010.

Gouyon P.h., Henry J.-P., Arnoud J., **Gene Avatars, The Neo-Darwinian Theory of Evolution**, Kluwer, 2002.

Griffiths, P.E., 2001, "Genetic Information: A Metaphor in Search of a Theory", **Philosophy of Science**, 68: 394–412.

Hanahan D, Weinberg R.A., "The hallmarks of cancer". **Cell,**100, 57–70, 2000.

Herrenschmidt Cl., **Les trois écritures**. Gallimard, 2007.

Huang S. "The war on cancer: lessons from the war on terror". **Front Oncol.** 4:293, 2014.

Huang S., Jaurand MC, Kamp D., Whysner J. Heil T., "Role of Mutagenicity in Asbestos Fiber-Induced Carcinogenicity and Other Diseases", <u>J Toxicol Environ Health B Crit Rev</u>. Jan-Jun; 14(1-4): 179–245, 2011.

Kato ., Scott M. Lippman, Keith T . Flaherty and Razelle Kurzrock, "The Conundrum of Genetic "Drivers" in Benign Conditions", **J Natl Cancer Inst,** 108 (8), 2016.

Karp RD, Johnson KH, Buoen LC, Ghobrial HK, Brand I, Brand KG. "Tumorigenesis by Millipore filters in mice: histology and ultrastructure of tissue reactions as related to pore size". **J Natl Cancer Inst.** 51(4):1275-85, 1973.

Kupiec, J.J., "A probabilistic theory for cell differentiation, embryonic mortality and DNA C-value paradox". **Specul Sci Technol** 6, 471-478, 1983.

Kupiec J.-J., "A Chance-Selection Model for Cellular Differentiation", **Cells, Death & Differentiation**, 3, 385-390, 1996.

Kuznetsov VA, Knott GD and Bonner RF: General statistics of stochastic process of gene expression in eukaryotic cells. Genetics, 161(3):1321–1332, 2002.

Jacob F., "Leçon inaugurale", **Collège de France**, 7 mai 1965.

Jaeger G., **Entanglement, information, and the interpretation of quantum mechanics,** Heildelberg: Springer, 2009.

Longo G. "Incompletezza*" per* **"La Matematica"**, vol. 4, Einaudi, 2010 (English version downloadable).

Longo G. "Reflections on Concrete Incompleteness," in **Philosophia Mathematica**, 19(3): 255-280, 2011.

Longo G. "Mathematical Infinity "in prospettiva" and the Spaces of Possibilities". *In* **Visible**, Semiotics J., n. 9, 2011b.

Longo G. "Science, Problem Solving and Bibliometrics". In **Use and Abuse of Bibliometrics**, Wim Blockmans et al. (eds), Portland Press, 2014.

Longo G., "The consequences of Philosophy", **Glass-Bead** Web Journal, http://www.glass-bead.org/article/the-consequences-of-philosophy/?lang=enview, 2015.

Longo G., "A review-essay on reductionism: some reasons for reading *"Reductionism, Emergence and Levels of Reality. The Importance of Being Borderline"*, a book by S. Chibbaro, L. Rondoni, A. Vulpiani. **Urbanomic**, London, https://www.urbanomic.com/document/on-the-borderline/ , May 8, 2016.

Longo G., "How Future Depends on Past Histories and Rare Events in Systems of Life", *to appear*, **Foundations of Science,** 2017 (versione preliminare in italiano in *Paradigmi*, n. XXXIII, Agosto, 2015, Gagliasso e Sterpetti eds).

Longo G., P. A. Miquel, C. Sonnenschein, A. Soto. "Is Information a proper observable for biological organization?" **Prog. Biophys. Mol. Biol.**, Vol. 109, Issue 3, pp. 108-114, August 2012.

Longo G., Montévil M.. **Perspectives on Organisms: Biological Time, Symmetries and Singularities.** Dordrecht: Springer, 2014.

Longo G., Montévil M., Sonnenschein C., Soto A.. "In Search of Principles for a Theory of Organisms", *in* **Journal of Biosciences**, Springer, pp. 955–968, 40(5), December, 2015.

Longo, G., Soto, A.M., "Why do we need theories?" **Prog. Biophys. Mol. Biol.**, 122, 4-10, Soto, Longo eds., 2016.

Maffini, M.V., Calabro, J.M., Soto, A.M., Sonnenschein, C., "Stromal regulation of neoplastic development: Age-dependent normalization of neoplastic mammary cells by mammary stroma". **Am. J. Pathol.** 167, 1405-1410, 2005.

Maltoni C, Lodi P, Masina A, *et al*. "Mesoteliomi negli operai di officine di grandi riparazioni (OGR) delle Ferrovie dello Stato italiane, esposti ad asbesto". Primo resoconto. **Acta Oncol**; 7: 159-86, 1986.

Marinov, G.K., Williams, B.A., McCue, K., Schroth, G.P., Gertz, J., Myers, R.M., Wold, B.J., "From single-cell to cell-pool transcriptomes: stochasticity in gene expression and RNA splicing". **Genome Res**. 24, 496-510, 2014.

Martincorena I. et al. (15 more authors) "High burden and pervasive positive selection of somatic mutations in normal human skin", Science, Vol. 348 no. 6237 pp. 880-886, may 2015.

Maynard-Smith J. "The idea of Information in Biology". **The Quarter Rev of Biology** 74: 495-400, 1999.

Microsoft http://news.microsoft.com/stories/computingcancer/ accessed on 31/10/2016.

Monod J., **Le Hasard et la Nécessité**, PUF, 1970.

Montévil M. Mossio M. "Closure of constraints in biological organisation", **Journal of Theoretical Biology**, vol. 372: 179-191, 2015.

Montévil M., M. Mossio, A. Pocheville, G. Longo., "Theoretical principles for biology: Variation", **Prog. Biophys.**

**Mol. Biol.**, 122, 36-50, Soto, Longo eds., 2016.

Montévil, M., Speroni, L., Sonnenschein, C., Soto A.M., "Modeling mammary organogenesis from biological first principles: cells and their physical constraints". **Prog. Biophys. Mol. Biol.**, 122, 58-69, Soto, Longo eds., 2016b.

Mossio, M., Montévil, M., Longo, G., 2016. Theoretical principles for biology: Organization. **Prog. Biophys. Mol. Biol.**, 122, 24-35, Soto, Longo eds., 2016.

Nicolis G., I. Prigogine. **Self-organization in non-equilibrium systems**. New York, Wiley, 1977.

Nowell PC. "The clonal evolution of tumor cell populations". **Science,** 194(4260):23–28, 1976.

Onuchic J., Luthey-Schulten Z., Wolynes P. "Theory of protein folding: The Energy Landscape Perspective", **Annual Review of Physical Chemistry**, Vol. 48: 545-600, 1997.

Paldi A., "Stochastic gene expression during cell differentiation: order from disorder?" **Cell Mol. Life Sci.,** 60, 1775-1779, 2003.

Paloma Alonso-Magdalen et al. **Molecular and Cellular Endocrinology** 355; 201–207, 2012.

Perret N., Longo G., "Reductionist perspectives and the notion of information", *in* (Soto, Longo, 2016).

Pauling L. "Schrödinger contribution to Chemistry and Biology", *in* **Schrödinger: Centenary Celebration of a Polymath** (Kilmister ed.) Cambridge U. P., 1987.

Perutz M.F. "E. Schrödinger's What is Life ? and molecular biology", *in* **Schrödinger: Centenary Celebration of a Polymath** (Kilmister ed.) Cambridge U. P., 1987.

Pisco O., Huang S., "Non-genetic cancer cell plasticity and therapy-induced stemness in tumour relapse: 'What does not kill me strengthens me' ", **British Journal of Cancer,** 112, 1725–1732, 2015.

Richard E. et al. (15 authors) "Single-Cell-Based Analysis Highlights a Surge in Cell-to-Cell Molecular Variability Preceding Irreversible Commitment in a Differentiation Proces". **Plos Biology,** December 27, 2016 http://dx.doi.org/10.1371/journal.pbio.1002585

Raj A, Peskin CS, Tranchina D, Vargas DY, Tyagi S., "Stochastic mRNA synthesis in mammalian cells". **PLoS Biol** 4:e309, 2006.

Raj A., R. van Oudenaarden. 'Stochastic Gene Expression and its Consequences'. **Cell**, 135(2): 216–226, 2008.

Rogers H., **Theory of Recursive Functions and Effective Computability,** McGraw Hill, 1967.

Smorinski C. "The incompleteness theorem" *in* **Handbook of Mathemaical Logic** (Barwais ed.), North Holland, 1978.

Smythies J. "Intercellular Signaling in Cancer—the SMT and TOFT Hypotheses, Exosomes, Telocytes and Metastases: Is the Messenger in the Message?" **Journal of Cancer**; 6(7): 604-609, 2015.

Sonnenschein C., Soto A.M. **The society of cells: cancer and control of cell proliferation**. Springer, 1999.

Sonnenschein C., Soto A. M. "The Death of the Cancer Cell", **Cancer Res.**; 71:4334-4337, 2011.

Sonnenschein C., Soto A. M. "The tissue organization field theory of cancer: A testable replacement for the somatic mutation theory", **Bioessays** 33: 332–340, 2011.

Sonnenschein C., Soto A. M. "The aging of the 2000 and 2011 hallmarks of cancer reviews: a critique". **Journal of Biosciences**; 38:651-63, 2013.

Sonnenschein C., Soto A. M. "Cancer Metastases: So Close and So Far". **J. Natl Cancer Inst,** 107(11), 2015.

Sonnenschein C, Davis B, Soto AM. "A novel pathogenic classification of cancer"s. **Cell Cancer Int**. 14: 113, 2014.

Soto A., Longo G. eds., *From the century of the genome to the century of the organism: New theoretical approaches*. Special issue, **Prog. Biophys. Mol. Biol.**, 122, 2016.

Soto, A.M., Longo, G., Montévil, M., Sonnenschein, C.,. "The biological default state of cell proliferation with variation and motility, a fundamental principle for a theory of organisms". **Prog. Biophys. Mol. Biol.,** 1 22, 16-23, Soto, Longo eds., 2016.

Soto, A.M., Longo, G., Miquel, P-A Montévil, M., Mossio, M., Perret, N., Pocheville, A., Sonnenschein, C.. "Toward a theory of organisms: Three founding principles in search of a useful integration". **Prog. Biophys. Mol. Biol.,** 122, 77-82, Soto, Longo eds., 2016.

Soto AM, Sonnenschein C. "Environmental causes of cancer: endocrine disruptors as carcinogens". **Nat Rev Endocrinol,** Jul;6(7):363-70, 2010.

Sporn MB., "The war on cancer" **Lancet** .1996;347(9012):1377–1381

Stanford Encyclopedia of Philosophy, accessed on 31/10/2016: http://plato.stanford.edu/entries/information-biological/

Straus, D.S. "Somatic mutation, cellular differentiation, and cancer causation". J. Natl. Cancer Inst. 67:233–241,1981.

Venter C., "We Have Learned Nothing from the Genome", **Der Spiegel**, July 29, 2010.

Versteeg R., "Tumors outside the mutation box", **Nature**, doi:10.1038/nature13061, vol. 1, 2014.

Vogelstein B, Papadopoulos N, Velculescu VE, Zhou S,Diaz Jr, LA, Kinzler KW. "Cancer genome landscapes". **Science**, 339(6127):546-58, 2013.

Weinberg R., **"**Coming Full Circle - form endless complexity to simplicity and back again**''**, **Cell 157,** March 27, 2014.

Wolfram S., "The importance of Unversal Computation", *in* **A. Turing, his work and impact**, Cooper ed.,Elsev., 2013.

Zhang XH, Jin X, Malladi S, et al. "Selection of bone metastasis seeds by mesenchymal signals in the primary tumor stroma". **Cell**.;154(5):1060–1073, 2013.